\documentclass[12pt]{article}
\usepackage{amsmath}
\usepackage{graphicx,psfrag,epsf}
\usepackage{enumerate}
\usepackage[round]{natbib}
\usepackage{url} % not crucial - just used below for the URL 
\usepackage{amsmath,amssymb,amsfonts,amsthm}
\usepackage{float,graphicx, multirow, setspace}
\usepackage{subfiles}
\usepackage{placeins}
\usepackage{url}
\usepackage{xr, xr-hyper}
\usepackage{hyperref}
\usepackage{algpseudocode}
\usepackage{algorithm}
\usepackage{subfig}

\usepackage[dvipsnames]{xcolor}
\newcommand{\blind}{0}

\begin{document}

\bibliographystyle{apalike}

\def\spacingset#1{\renewcommand{\baselinestretch}%
{#1}\small\normalsize} \spacingset{1}

%%%%%%%%%%%%%%%%%%%%%%%%%%%%%%%%%%%%%%%%%%%%%%%%%%%%%%%%%%%%%%%%%%%%%%%%%%%%%%

\newcommand{\inserttitle}{\bf{
  Putting Skill as Nearly Indistinguishable from Noise: An Empirical Bayes Analysis of PGA Tour Performance
  % Putting for Applause, Not for Odds: An empirical Bayes Comparison of Putting, Approaching, and Driving Skills in Golfff
}}

\if0\blind
{
  \title{
    \inserttitle
  }

  \author{
    Ryan S. Brill\thanks{
      Graduate Group in Applied Mathematics and Computational Science, University of Pennsylvania. Correspondence to: ryguy123@sas.upenn.edu
    } \ 
    and Abraham J. Wyner\thanks{
      Dept.~of Statistics and Data Science, The Wharton School, University of Pennsylvania
    }
  }
  \maketitle
} \fi

\if1\blind
{
  \bigskip
  \bigskip
  \bigskip
  \begin{center}
  \title{
    \inserttitle
  }
  \end{center}
  \medskip
} \fi

\bigskip
\begin{abstract}
We revisit a foundational question in golf analytics: how important are the core components of performance––driving, approach play, and putting––in explaining success on the PGA Tour? Building on Mark Broadie's strokes gained analyses, we use an empirical Bayes approach to estimate latent golfer skill and assess statistical significance using a multiple testing procedure that controls the false discovery rate. While tee-to-green skill shows clear and substantial differences across players, putting skill is both less variable and far less reliably estimable. Indeed, putting performance appears nearly indistinguishable from noise.

% We use an empirical Bayes framework to estimate PGA Tour golfers' skill across driving, approach play, and putting using shot-level strokes gained data from one PGA Tour season. While tee-to-green skill shows clear and significant differences across players, putting skill appears both less variable and far less reliably estimable. Applying a multiple testing procedure that controls the false discovery rate, we find that putting performance is nearly indistinguishable from noise.

\end{abstract}

\noindent%

\vfill
\newpage
\spacingset{1.45} % DON'T change the spacing!

%%%%%%%%%%%%%%%%%%%%%%%%%%%%%%%%%%%%%%%%%%%%%%%%%%%%%
%%%%%%%%%%%%%%%%%%%%%%%%%%%%%%%%%%%%%%%%%%%%%%%%%%%%%
%%%%%%%%%%%%%%%%%%%%%%%%%%%%%%%%%%%%%%%%%%%%%%%%%%%%%

%%%%%%%%%%%%%%%%%%%%%%%%%%%%%%%%%%%%%%%%%%%%%%%%%%%%%
\section{Introduction}\label{sec:intro}

``Drive for show, putt for dough'' is a well-known and long-standing adage in golf. It contains some truth––how well a golfer putts during a tournament is predictive of outcomes like winning and earnings \citep{datagolf2017puttingvsballstriking,jensen2023puttfordough}. But looking forward to future tournaments, the picture changes. In a seminal series of works, \citet{broadie08,MarkBroadieStrokesGained} introduced \textit{strokes gained}––which quantifies the quality of a shot relative to an average PGA Tour golfer––demonstrating that season-long success hinges more on superior driving and approach play than on putting. This conclusion has been reinforced by subsequent analyses (e.g., \citet{datagolf2017puttingvsballstriking,floyd2020puttingrandomness}). Elite players consistently gain more strokes on the field tee-to-green than on the green, dispelling the notion that putting is the most decisive component of a golfer's underlying skill.

Prior analyses highlighting the relative importance of ball-striking over putting have focused on decomposing variation in strokes gained across shot types without addressing questions of statistical significance or uncertainty. As a result, we are left to wonder: do observed differences in estimated skill reflect true differences in ability, or are they merely artifacts of random variation? In this paper, we revisit and extend these findings using statistical methods that are well suited to the structure of strokes gained data––methods that not only provide skill estimates but also allow us to rigorously distinguish signal from noise.

Our analysis focuses on strokes gained data from the 2015 and 2017 PGA Tour seasons, comprising shot-level outcomes for over 550 golfer-seasons across 83 tournaments.
Central to our analysis is the use of empirical Bayes to estimate golfers' underlying strengths across putting, approach play, and driving. While researchers have previously applied empirical Bayes and other sophisticated statistical modeling techniques to various golf performance problems (e.g., \citet{baker2016rydercup,connolly2008pgatour,datagolf2022masters}), we view our fully open-source empirical Bayes analysis of strokes gained data for estimating golfer skill as a welcome addition to the literature. We believe this approach is especially well suited to the task. Its Bayesian structure provides principled adjustment for varying sample sizes––shrinking estimates more heavily toward the mean for golfers with less data––while its empirical formulation avoids the computational burden of MCMC or other sampling algorithms. The result is a fast, data-driven estimation framework that produces not only skill estimates, but also a full posterior distribution, enabling assessments of uncertainty and significance.

Using this framework, we corroborate Broadie's central claim and then assess the significance of observed differences using simultaneous inference across players within each shot type. Applying the Benjamini–Hochberg procedure to control the false discovery rate, we find that variation in putting skill is both markedly less dispersed and less reliably estimable than variation in tee-to-green skill. Indeed, across players, putting skill appears nearly indistinguishable from noise. 
Our analysis thus provides further statistical evidence––grounded in both estimation and inference––for the claim that putting, despite its volatility and dramatic importance in individual tournaments, is far less informative of a golfer's true ability than driving and approach play.

The remainder of this article is organized as follows. In Section \ref{sec:data} we describe our dataset. In Section~\ref{sec:emp_bayes} we use empirical Bayes to estimate golfers' underlying strengths and in Section~\ref{sec:significance} we evaluate the statistical significance of these estimates using multiple testing procedures. We conclude in Section~\ref{sec:discussion}.

%%%%%%%%%%%%%%%%%%%%%%%%%%%%%%%%%%%%%%%%%%%%%%%%%%%%%
\section{Data}\label{sec:data}

We obtained a stroke-by-stroke dataset from the 2015 and 2017 PGA Tour seasons. 
For each recorded shot, the dataset includes the golfer's identity, shot type (which we categorize into three bins: driving, approaching, and putting), and the strokes gained value––a performance metric that quantifies the quality of a shot relative to a baseline average PGA Tour golfer \citep{broadie08,MarkBroadieStrokesGained}. 
Specifically, strokes gained measures how much a golfer improves (or worsens) his position compared to the expected number of strokes it would take a tour-average player to complete the hole from the same starting location.

\begin{figure}[hbt!]
    \centering{}
    \includegraphics[width=0.5\textwidth]{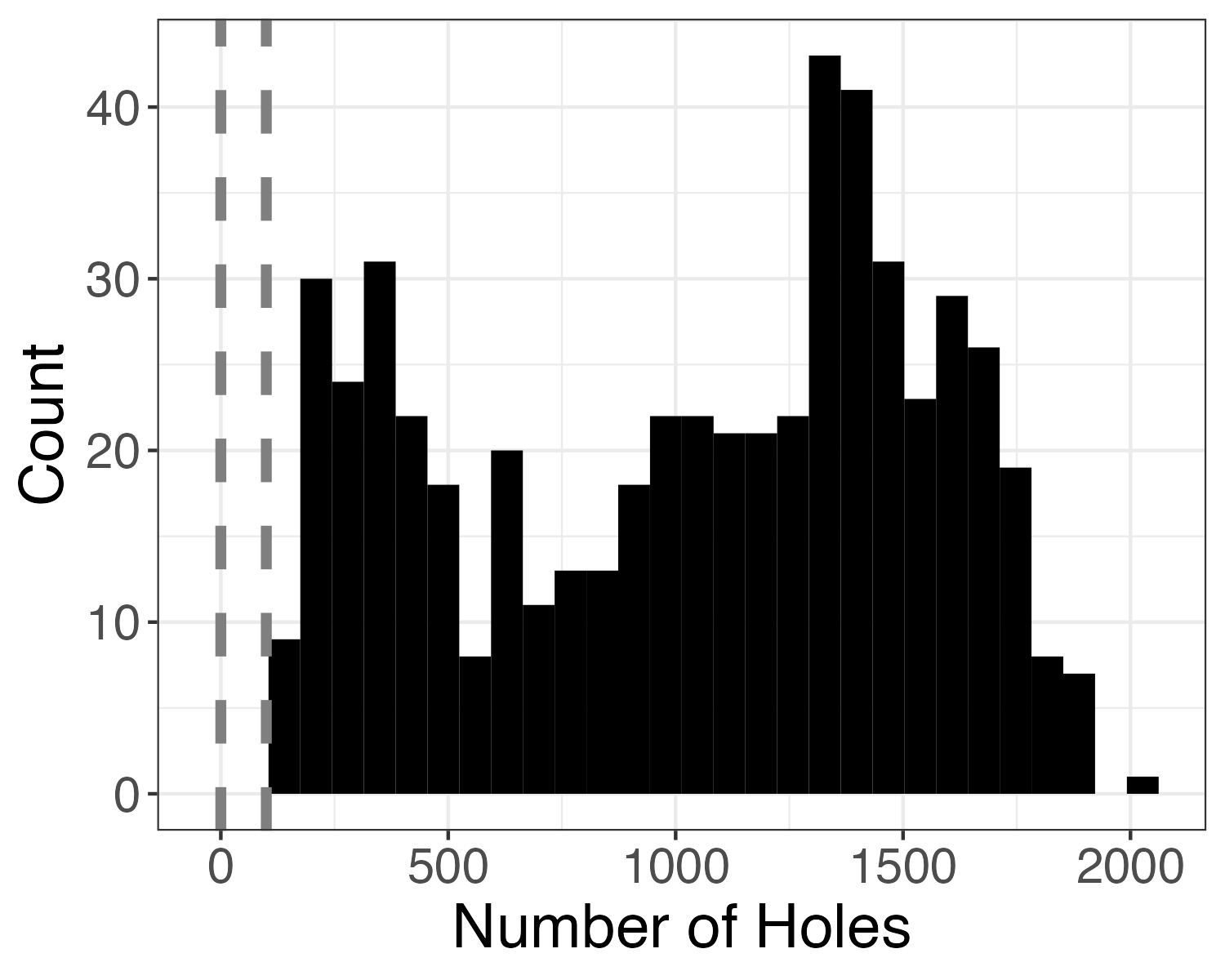}
    \caption{
        The distribution of the number of holes played per golfer-season.
    }
    \label{fig:plot_nHoles_hist}
\end{figure}

\begin{figure}[hbt!]
    \centering{}
    \includegraphics[width=1\textwidth]{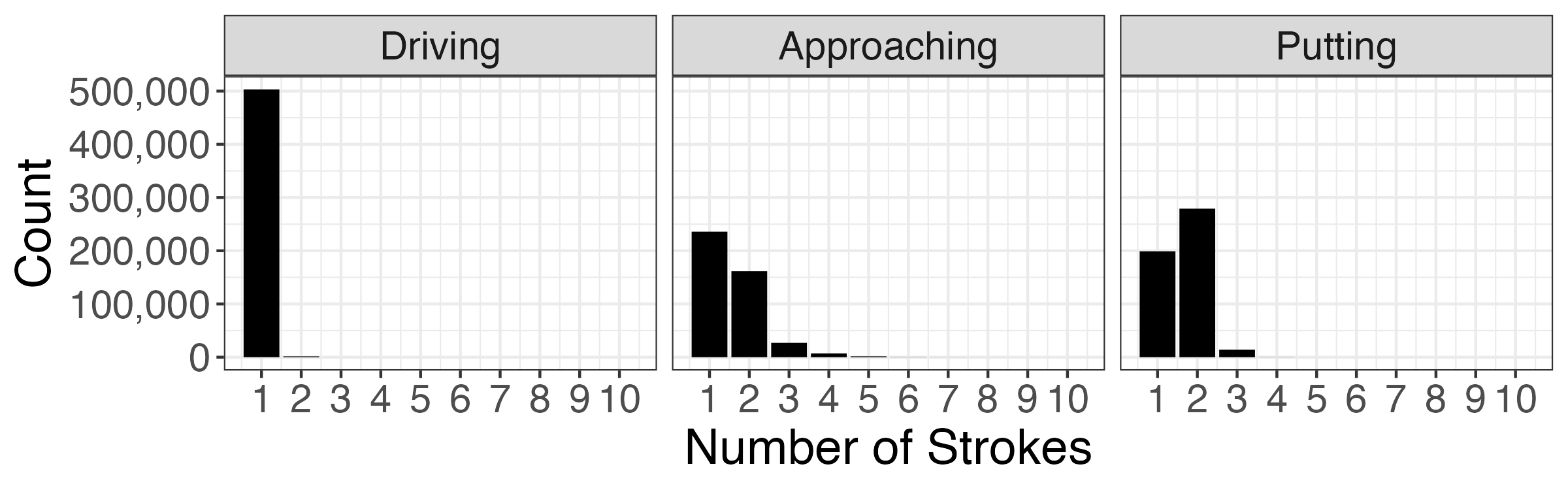}
    \caption{
        The distribution of the number of strokes per hole by stroke category.
    }
    \label{fig:plot_Nstrokes_hist}
\end{figure}

After filtering out golfer-seasons who played fewer than $150$ holes, the final cohort consists of $n=553$ golfer-seasons. 
The cleaned dataset includes 83 tournaments and comprises 504,742 player-holes, 504,742 player-drives, 492,291 player-putts, and 433,399 player-approach shots. Figure~\ref{fig:plot_nHoles_hist} displays the distribution of the number of holes played per golfer-season. Figure~\ref{fig:plot_Nstrokes_hist} shows the distribution of the number of strokes per hole by stroke category.

Our code is publicly available on GitHub.\footnote{\url{https://github.com/snoopryan123/golf_empirical_bayes_skill}} Unfortunately, due to an NDA, we are unable to include the dataset on GitHub.

%%%%%%%%%%%%%%%%%%
%%%%%%%%%%%%%%%%%%

%%%%%%%%%%%%%%%%%%
%%%%%%%%%%%%%%%%%%

%%%%%%%%%%%%%%%%%%%%%%%%%%%%%%%%%%%%%%%%%%%%%%%%%%%%%
\section{Empirical Bayes estimation of golfer skill}\label{sec:emp_bayes}

%%%%%%%%%%%%%%%%%%%%%%%%%%%%%%%%%%%%%%%%%%%%%%%%%%%%%
\subsection{Model specification}

We categorize strokes into three types: driving, approaching, and putting. For each golfer-season and stroke category, we estimate a latent skill parameter using an empirical Bayes approach. The model and estimation technique are analogous across the three stroke categories. 
We separate players by season, treating different years of the same player as distinct, which ensures compatibility with the empirical Bayes approach.

Let $i=1,...,n$ index the golfer-seasons in our dataset, and let $s \in \{\text{driving}, \text{approaching}, \text{putting}\}$ denote the stroke category. For golfer $i$, let $j=1,...,N_{is}$ index the holes for which we observe strokes in category $s$.
Define $X_{ijs}$ as the observed outcome for golfer $i$ on hole $j$ in stroke category $s$.
\begin{itemize}
    \item For \textit{driving}, $X_{ijs}$ is the strokes gained on the drive taken on hole $j$.
    \item For \textit{putting}, $X_{ijs}$ is the number of putts under expected on hole $j$, equivalently the sum of strokes gained across all putts on that hole.
    \item For \textit{approaching}, $X_{ijs}$ is the number of strokes under expected on hole $j$, equivalently the sum of strokes gained across all approach shots taken on hole $j$.
\end{itemize}
In all cases, a higher value of $X_{ijs}$ reflects better performance on that hole for the corresponding stroke category.

We model the observed outcomes using a two-level Gaussian hierarchical model:
\begin{align}
\begin{split}
\label{eqn:emp_bayes}
    X_{ijs} &\sim \mathcal{N}(\mu_{is}, \ \sigma^2_{is}), \\
    \mu_{is} &\sim \mathcal{N}(\mu_s, \ \tau^2_s).
\end{split}
\end{align}

Here, $\mu_{is}$ represents golfer $i$'s unobserved latent skill in stroke category $s$, while $\sigma^2_{is}$ captures the variance in performance across holes for that golfer and stroke type. The hyperparameters $\mu_s$ and $\tau^2_s$ represent the population-level mean and variance in latent skill across all golfers for stroke category $s$.

\begin{figure}[hbt!]
    \centering
    \includegraphics[width=1\textwidth]{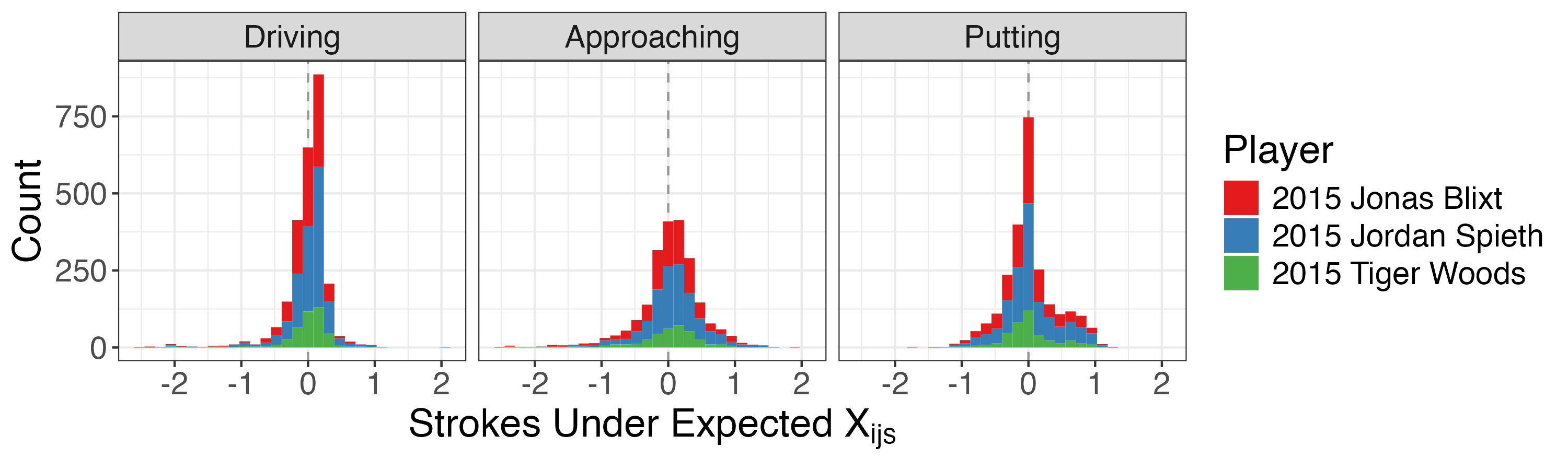}
    \caption{
        Empirical distributions of hole-level stroke outcomes $\{X_{ijs} : j = 1,...,N_{is}\}$ for three representative golfers $i$ (color) and each stroke category $s$ (facet).
    }
    \label{fig:plot_exGolferDists}
\end{figure}

In Figure~\ref{fig:plot_exGolferDists}, we visualize the distribution of hole-level stroke outcomes (i.e., total strokes gained) for three representative golfers, separated by stroke category. 
Although the observed outcomes $X_{ijs}$ are not strictly normally distributed, the Gaussian assumption provides a reasonable approximation. 
We adopt this model not because it precisely captures the distributional shape of stroke-level data, but because it yields interpretable and effective estimators of latent skill.

%%%%%%%%%%%%%%%%%%
%%%%%%%%%%%%%%%%%%

We estimate golfer $i$'s quality for stroke category $s$, denoted $\mu_{is}$, using the posterior mean. Under our Normal-Normal conjugate model~\eqref{eqn:emp_bayes}, this posterior mean has a closed-form expression:
\begin{align}
\begin{split}
\label{eqn:post_mean}
    \widehat{\mu}_{is} := \mathbb{E}[\mu_{is} \mid X_{i1s}, \ldots, X_{iN_{is}s}] = \frac{\frac{1}{\sigma^2_{is}} \sum_{j=1}^{N_{is}} X_{ijs} + \frac{\mu_s}{\tau^2_s}}{\frac{N_{is}}{\sigma^2_{is}} + \frac{1}{\tau^2_s}}.
\end{split}
\end{align}
%%%%%
This posterior mean is a weighted average of the golfer's observed mean performance and the population-level prior mean $\mu_s$. The weights depend on the number of holes played $N_{is}$, the golfer's performance variance $\sigma^2_{is}$, and the prior variance $\tau^2_s$.
%%%%%
As $N_{is}$ increases, the influence of the observed data grows, and the posterior mean $\widehat{\mu}_{is}$ approaches the sample mean $\overline{X}_{is}$. Conversely, for golfers with limited data, the estimate shrinks more heavily toward the prior mean $\mu_s$. This adaptivity reflects the Bayesian principle of balancing individual-level evidence with population-level regularization.

Estimator~\eqref{eqn:post_mean} depends on the unknown parameters $\mu_s$, $\tau^2_s$, and $\sigma^2_{is}$. To apply this estimator in practice, we adopt an \textit{empirical Bayes} (EB) approach \citep{brown2008,GridWAR}. Specifically, we replace the unknown parameters in Equation~\eqref{eqn:post_mean} with their maximum likelihood estimates (MLEs).
Let $\widehat{\mu}_s$, $\widehat{\tau}^2_s$, and $\widehat{\sigma}^2_{is}$ denote the MLEs of $\mu_s$, $\tau^2_s$, and $\sigma^2_{is}$, respectively. We compute these estimates using the iterative procedure described in Algorithm 1 of Appendix C.1 in \citet{GridWAR}. These plug-in estimates yield a fully data-driven empirical Bayes estimator of latent golfer skill,
\begin{equation}
\label{eqn:mu_hat_EB}
\widehat{\mu}_{is}^{(\text{EB})} := \frac{\frac{1}{\widehat{\sigma}^2_{is}} \sum_{j=1}^{N_{is}} X_{ijs} + \frac{\widehat{\mu}_s}{\widehat{\tau}^2_s}}{\frac{N_{is}}{\widehat{\sigma}^2_{is}} + \frac{1}{\widehat{\tau}^2_s}}.
\end{equation}

%%%%%%%%%%%%%%%%%%%%%%%%%%%%%%%%%%%%%%%%%%%%%%%%%%%%%
\subsection{Results}\label{sec:results}

\begin{figure}[hbt!]
    \centering
    \includegraphics[width=1\textwidth]{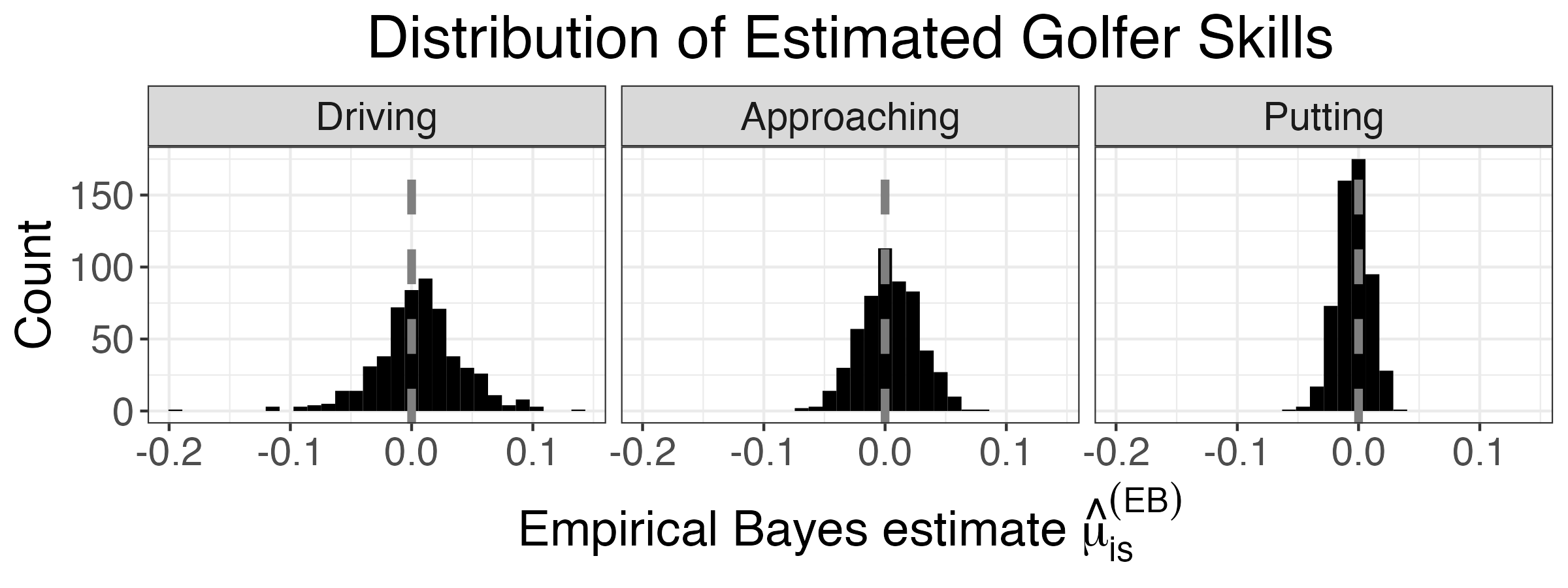}
    \caption{
        Distribution of estimated golfer skill values $\{ \widehat{\mu}_{is}^{(\text{EB})} \}_{i=1}^{n}$ by stroke category. 
    }
    \label{fig:plotEB}
\end{figure}

In Figure~\ref{fig:plotEB}, we visualize the distribution of estimated golfer skill values $\{ \widehat{\mu}_{is}^{(\text{EB})} \}_{i=1}^{n}$ for each stroke category $s$. Recall that the latent skill parameter $\mu_{is}$ is expressed on the scale of 
% strokes gained per hole, 
strokes under expected per hole, 
where higher values indicate better performance.
The variability in estimated skill is substantially smaller for putting than approaching, and is smaller for approaching than driving.
This suggests that the gap between the best and worst putters is narrower than the gap in other stroke types. 
%%%%%
To contextualize the effect sizes, note that the difference in skill between the $95^\text{th}$ and $5^\text{th}$ percentile golfers is $\Delta_\text{driving} = 0.115$, $\Delta_\text{approaching} = 0.081$, and $\Delta_\text{putting} = 0.043$ strokes gained per hole. 
Cumulatively, over a 72-hole tournament, these differences amount to $8.3$ strokes gained in driving, $5.9$ strokes gained in approaching, and $3.1$ strokes gained in putting, together representing a competitive edge exceeding $17$ strokes across the event.
While differences in putting skill are meaningful, they account for much less than one-third of the total skill gap and are therefore substantially less predictive of tournament performance compared to tee-to-green skills like driving and approach play.

%%%%%%%%%%%%%%%%%%
%%%%%%%%%%%%%%%%%%

\begin{figure}[hbt!]
    \centering
    \includegraphics[width=1\textwidth]{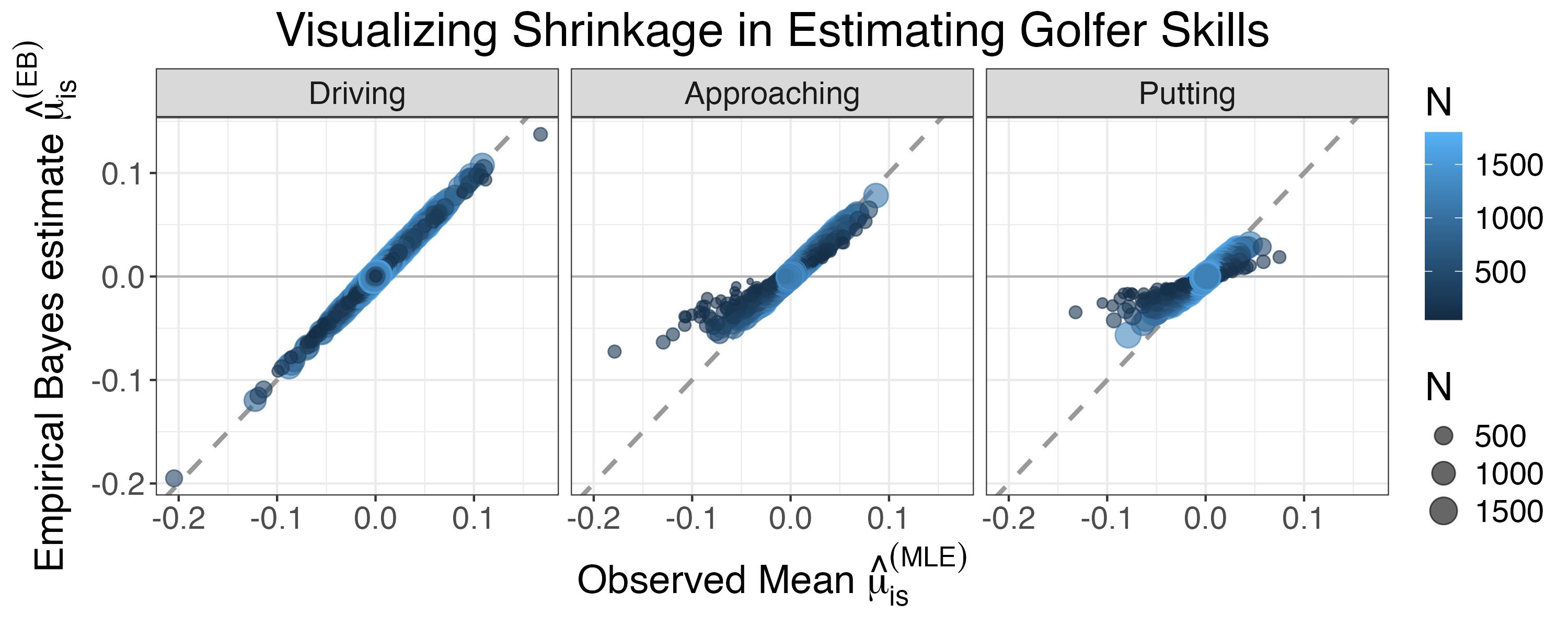}
    \caption{
        For each golfer $i$ and stroke category $s$ (facet), we plot the estimated skill $\widehat{\mu}_{is}^{(\text{EB})}$ (posterior mean, $y$-axis) against the observed mean hole-level outcome $\widehat{\mu}_{is}^{(\text{MLE})} = \overline{X}_{is}$ (MLE, $x$-axis).
        Points are colored and sized according to the number of holes $N_{is}$ (blue = large $N_{is}$, gray = small $N_{is}$).
        The dashed diagonal line indicates the identity line $y = x$, along which the empirical Bayes estimator equals the MLE.
    }
    \label{fig:plotEBshrinkage}
\end{figure}

In Figure~\ref{fig:plotEBshrinkage}, we visualize the extent of shrinkage in the empirical Bayes estimators. For each stroke category $s$ and golfer $i$, we plot the estimated skill $\widehat{\mu}_{is}^{(\text{EB})}$ (posterior mean) against the golfer's empirical mean hole-level outcome (i.e., the MLE $\widehat{\mu}_{is}^{(\text{MLE})} = \overline{X}_{is}$). 
%%%%%
Each point is colored and sized according to the number of holes $N_{is}$: small gray dots represent golfers with limited data, while large blue dots correspond to golfers with substantial hole counts. For high-sample golfers, the empirical Bayes estimator $\widehat{\mu}_{is}^{(\text{EB})}$ lies close to the identity line ($y = x$), reflecting minimal shrinkage. In contrast, low-sample golfers exhibit significant shrinkage toward the overall mean skill $\widehat{\mu}_s \approx 0$.
%%%%%
The degree of shrinkage differs across stroke categories. We observe the most aggressive shrinkage in putting, followed by approach shots, with driving skill exhibiting the least shrinkage. This pattern is consistent with the lower signal-to-noise ratio and tighter skill distribution in putting compared to other stroke types.

%%%%%%%%%%%%%%%%%%
%%%%%%%%%%%%%%%%%%

\begin{figure}[p!]
    \centering
    \includegraphics[width=0.9\textwidth]{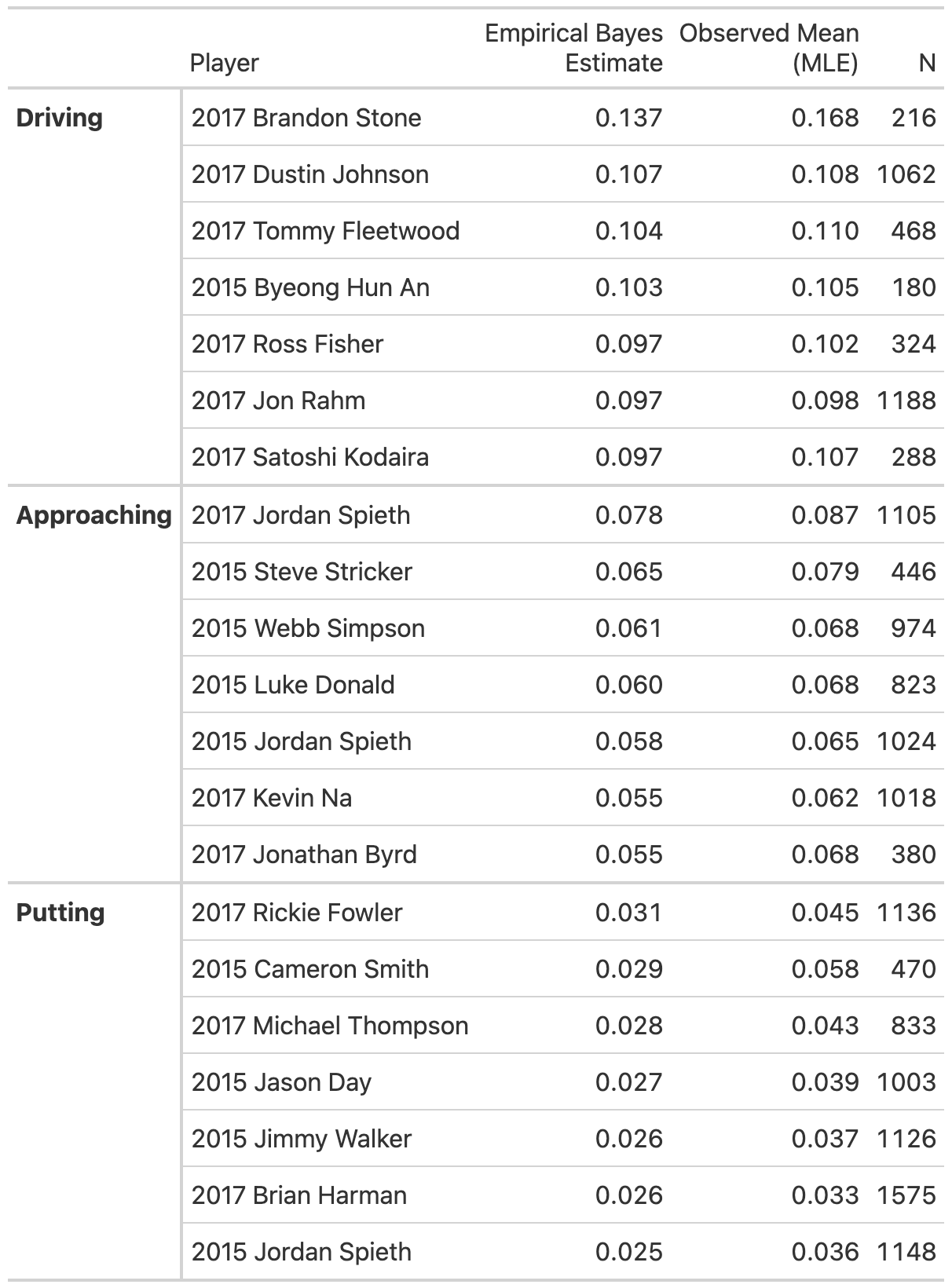}
    \caption{
        The empirical Bayes estimate $\widehat{\mu}_{is}^{(\text{EB})}$, observed mean $\widehat{\mu}_{is}^{(\text{MLE})}$, and number of holes $N_{is}$ for the top seven golfers in each stroke category.
    }
    \label{fig:plotTopGolfers}
\end{figure}

In Figure~\ref{fig:plotTopGolfers}, we display the estimated skill values $\widehat{\mu}_{is}^{(\text{EB})}$ alongside the number of holes $N_{is}$ for the top seven performers in each stroke category.
2015 Jordan Spieth appears in the top seven for both approaching and putting skill. Widely considered his breakout season, Spieth's 2015 campaign included five PGA Tour victories—most notably, wins at The Masters and the U.S. Open. He also finished in the top three in all four majors and earned both the PGA Tour Player of the Year and PGA Player of the Year honors.\footnote{
  \href{https://www.todays-golfer.com/news-and-events/tour-news/jordan-spieth-majors-comeback-2025/}{https://www.todays-golfer.com/news-and-events/tour-news/jordan-spieth-majors-comeback-2025/}
}
% %%%%%
% Similarly, Jason Day ranks among the top seven in both driving and putting skill. His breakout year in 2015 included five PGA Tour wins, highlighted by his victory at the PGA Championship. Day's success was underscored by his exceptional putting performance.\footnote{
%   \href{https://www.cbssports.com/golf/news/jason-days-2015-16-putting-was-the-best-recorded-performance-in-recent-history/}{https://www.cbssports.com/golf/news/jason-days-2015-16-putting-was-the-best-recorded-performance-in-recent-history/}
% }

%%%%%%%%%%%%%%%%%%
% \begin{figure}[hbt!]
%%%%%%%%%%%%%%%%%%

%%%%%%%%%%%%%%%%%%%%%%%%%%%%%%%%%%%%%%%%%%%%%%%%%%%%%
\section{Significance}\label{sec:significance}

%%%%%%%%%%%%%%%%%%%%%%%%%%%%%%%%%%%%%%%%%%%%%%%%%%%%%
\subsection{Methodology}

In Section~\ref{sec:results}, we observed that the distribution of putting skill is considerably tighter and more heavily shrunk toward the mean than the distributions for approaching and driving skill. This raises a natural question: are extreme skill values––particularly in putting––statistically significant? More broadly, how confident can we be that observed differences in estimated skill reflect genuine differences in ability rather than fluctuations attributable to noise?
Hence, in this section, we assess the significance of golfer skill estimates across stroke categories.

For each golfer $i$ and stroke category $s$, we test whether the latent skill parameter is significantly different from zero:
\begin{equation}
H_0: \mu_{is} = 0 \quad \text{versus} \quad H_1: \mu_{is} \neq 0.
\end{equation}
From our Bayesian model in Equation~\eqref{eqn:emp_bayes}, the posterior distribution of $\mu_{is}$ given the observed data is
\begin{equation}
\label{eqn:post_dist}
\mu_{is} \mid X_{i1s}, \ldots, X_{iN_{is}s} \sim \mathcal{N}(\widehat{\mu}_{is},\ \widehat{V}_{is}),
\end{equation}
where $\widehat{\mu}_{is}$ is the posterior mean from Equation~\eqref{eqn:post_mean} and the posterior variance is
\begin{equation}
\label{eqn:post_var}
\widehat{V}_{is} = \left( \frac{N_{is}}{\sigma^2_{is}} + \frac{1}{\tau^2_s} \right)^{-1}.
\end{equation}
As discussed in Section~\ref{sec:results}, we employ an empirical Bayes approach and plug in the maximum likelihood estimates of the hyperparameters $\mu_s$, $\tau^2_s$, and $\sigma^2_{is}$ when computing $\widehat{\mu}_{is}$ and $\widehat{V}_{is}$.

Under the null hypothesis, the posterior probability that a golfer's skill $\mu_{is}$ is as or more extreme than the observed estimate $\widehat{\mu}_{is}$ is given by the two-sided Bayesian $p$-value:
\begin{equation}
\label{eqn:p_val}
p_{is} = 2 \cdot \Phi\left(\frac{-|\widehat{\mu}_{is}|}{\sqrt{\widehat{V}_{is}}}\right),
\end{equation}
where $\Phi$ denotes the standard Normal cumulative distribution function. A larger magnitude of the test statistic $|\widehat{\mu}_{is}|$ corresponds to a smaller $p$-value $p_{is}$, reflecting greater statistical significance.

For each stroke category $s$, we aim to identify the golfers $i$ for whom there is strong evidence that $\mu_{is} \neq 0$. This entails conducting $n = 553$ hypothesis tests per category. If all $n$ null hypotheses were true and we tested at the $\alpha = 0.05$ level, we would expect $n \cdot \alpha = 15.5$ false positives purely by chance.
To mitigate this issue and reduce the number of spurious discoveries, we apply the Benjamini--Hochberg procedure to control the \textit{False Discovery Rate} (FDR) \citep{benjaminiHochberg}. This method ensures that, at a chosen significance level $\alpha$, the expected proportion of false positives among the rejected null hypotheses is no greater than $\alpha$.

While our p-values are computed using empirical Bayes posteriors with plug-in estimates of hyperparameters––introducing a minor deviation from the strict uniformity assumption of $p$-values under the null––this effect is typically small and should not materially affect the validity of the Benjamini-Hochberg procedure \citep{Efron_2010,10.1111/j.1467-9868.2004.00439.x}. 
% Indeed, as shown in the putting panel of Figure~\ref{fig:plotBH}, the distribution of p-values is visually close to uniform, further supporting the reliability of the procedure under our modeling assumptions.

%%%%%%%%%%%%%%%%%%%%%%%%%%%%%%%%%%%%%%%%%%%%%%%%%%%%%
\subsection{Results}

\begin{figure}[hbt!]
    \centering
    \includegraphics[width=0.45\textwidth]{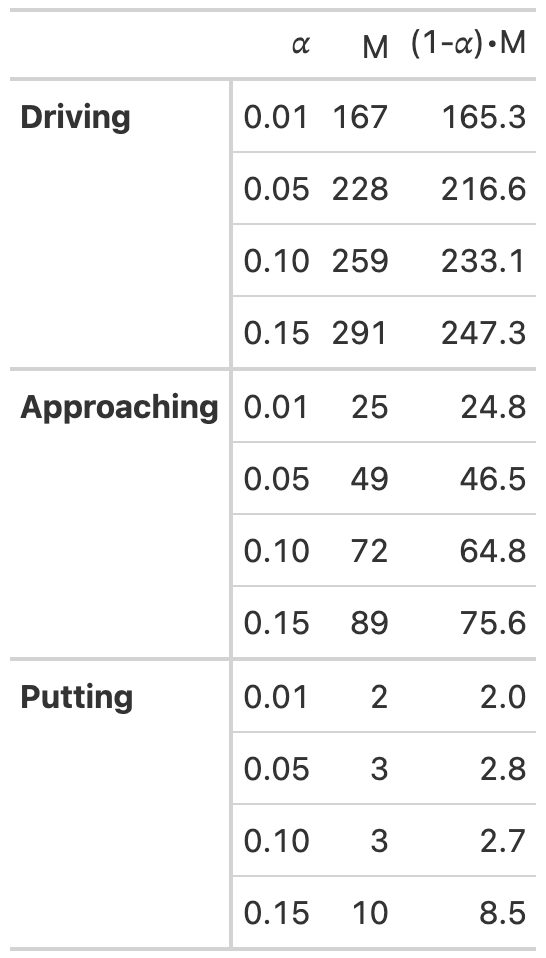}
    \caption{
        Number of statistically significant golfer skills identified by the Benjamini-Hochberg procedure for each stroke category across various significance levels $\alpha$. For each $\alpha$, we plot the total number of discoveries $M$ and the expected number of true discoveries $(1 - \alpha)\cdot M$.
    }
    \label{fig:plot_BH_nSig}
\end{figure}

In Figure~\ref{fig:plot_BH_nSig}, we present the results of the Benjamini--Hochberg procedure. For each stroke category and a range of significance levels $\alpha$, we report two quantities: the number of discoveries $M$ (i.e., the number of golfers identified with statistically significant skill) and the expected number of true discoveries, given by $(1 - \alpha) \cdot M$. This latter quantity reflects our control of the False Discovery Rate, which ensures that at most an $\alpha$ proportion of the $M$ discoveries are expected to be false positives. 

%%%%%%%%%%%%%%%%%%
%%%%%%%%%%%%%%%%%%

Just two of the 553 putters at significance level $0.01$––and a mere three at levels $0.05$ and $0.10$––are identified as having significantly nonzero skill. Even at the more lenient level of $0.15$, just ten putters are deemed significant. While we expect only one to two of these ten to be false positives under FDR control, we do not know which putters. %one it is. 
This underscores the remarkable uniformity of putting skill: the distribution is so tight that it is nearly indistinguishable from random noise at the individual level.

In contrast, the other stroke categories show much stronger signals. At the $\alpha = 0.10$ level, we identify 72 significant approachers and 259 significant drivers. In general, the number of significantly skilled drivers is roughly triple that of approachers, suggesting greater dispersion––and thus greater opportunity for separation––in driving skill.
% \fixme{Further commentary?}

% This pattern highlights a broader point: **the ability to distinguish individual skill depends not only on effect size but also on signal-to-noise ratio.** Putting appears to be both less variable and more prone to noise, making it harder to statistically separate elite performers from the rest. Conversely, larger performance gaps in driving and approach shots make individual excellence easier to detect.

\begin{figure}[hbt!]
    \centering
    \includegraphics[width=1\textwidth]{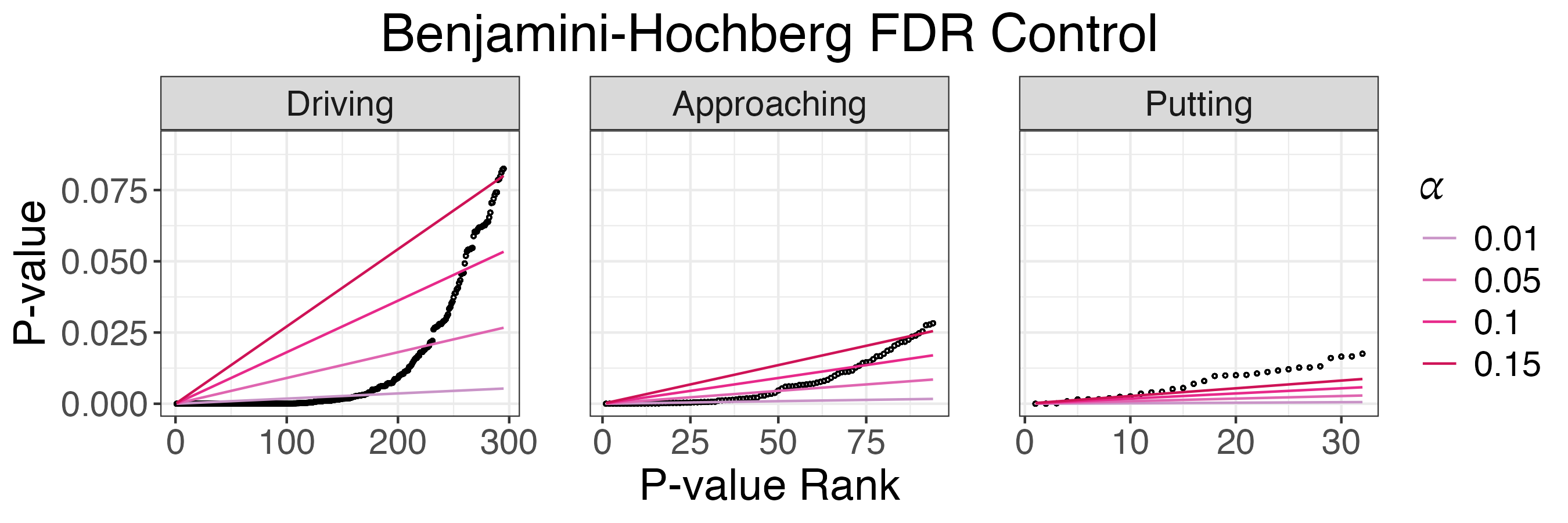}
    \caption{
        Visualization of the Benjamini-Hochberg procedure. Each black dot represents a hypothesis test, with $p$-value on the $y$-axis and its rank on the $x$-axis. Colored lines correspond to different significance levels $\alpha$; each line originates from the origin and has slope $\alpha/n$, where $n = 553$ is the number of golfers. For a given $\alpha$, the BH procedure rejects all null hypotheses with $p$-values below the rightmost intersection point between the line and the $p$-value curve.
    }
    \label{fig:plotBH}
\end{figure}

We further visualize the Benjamini-Hochberg (BH) procedure in Figure~\ref{fig:plotBH}. For each stroke category, we plot the $p$-values against their ranks and overlay multiple BH decision thresholds for varying values of $\alpha$.
For putters, the vast majority of $p$-value dots lie above all of the BH lines, indicating that almost no putting skills are statistically distinguishable from zero––even at higher significance levels. In contrast, many driver $p$-values fall below the BH lines, resulting in a substantial number of significant discoveries across all $\alpha$ levels. The case for approach shots lies in between: their $p$-values straddle the BH lines, making the number of significant approachers more sensitive to the choice of $\alpha$.

%%%%%%%%%%%%%%%%%%
%%%%%%%%%%%%%%%%%%

%%%%%%%%%%%%%%%%%%%%%%%%%%%%%%%%%%%%%%%%%%%%%%%%%%%%%
\section{Discussion}\label{sec:discussion}

This paper revisits a foundational question in golf analytics: how important are the core components of performance––driving, approach play, and putting––in explaining success on the PGA Tour? Further, how reliably can we estimate skill in each of these components? Building on previous strokes gained analyses, we use an empirical Bayes methodology to estimate latent golfer skill and assess statistical significance through a simultaneous inference procedure.

Consistent with prior work, we find that skill in driving and approach play is more consequential than putting: the spread in estimated skill is wider tee-to-green, and these components account for a larger share of cumulative performance differentials over a tournament. But beyond reaffirming this asymmetry, our analysis reveals a deeper insight: putting skill is not only less impactful, but also far less reliably estimable. Even with a full season of data, the variation in putting performance is so tight and noisy that only a handful of golfers exhibit statistically significant putting skill—compared to dozens in driving and approach play. This discrepancy reflects both a narrower distribution of putting ability and a lower signal-to-noise ratio, leading to greater shrinkage and fewer discoveries. The results suggest that much of what is perceived as putting prowess may reflect randomness rather than persistent ability, reinforcing the view that putting outcomes—though often dramatic—are relatively poor indicators of true skill.

More broadly, this work illustrates the value of empirical Bayes methods in sports analytics: they deliver skill estimates that are not only data-efficient and computationally scalable, but also naturally capture uncertainty and sample size heterogeneity. In addition, by enabling principled multiple hypothesis testing, they offer a rigorous framework for distinguishing true skill from noise across large pools of athletes. This framework may prove equally valuable in other sports contexts where performance is observed through noisy, high-frequency data.

% %%%%%%%%%%%%%%%%%%%%%%%%%%%%%%%%%%%%%%%%%%%%%%%%%%%%%
% %%%%%%%%%%%%%%%%%%%%%%%%%%%%%%%%%%%%%%%%%%%%%%%%%%%%%
% %%%%%%%%%%%%%%%%%%%%%%%%%%%%%%%%%%%%%%%%%%%%%%%%%%%%%
% \if0\blind
% {
%   \section*{Acknowledgments}
%   \input{acknowledgements}
% } \fi

%%%% REFERENCES
\bibliography{refs}

% %%%%%%%%%%%%%%%%%%%%%%%%%%%%%%%%%%%%%%%%%%%%%%%%%%%%%
% %%%%%%%%%%%%%%%%%%%%%%%%%%%%%%%%%%%%%%%%%%%%%%%%%%%%%
% %%%%%%%%%%%%%%%%%%%%%%%%%%%%%%%%%%%%%%%%%%%%%%%%%%%%%
% \clearpage
% \newpage
% % \bigskip
% \begin{center}
% {\large\bf SUPPLEMENTARY MATERIAL}
% \end{center}
% \appendix

\end{document}